%% file: cheatsheet_aka_2g_3g_4g_5g.tex
\documentclass[
10pt, % Main document font size
a4paper, % Paper type, use 'letterpaper' for US Letter paper
oneside, % One page layout (no page indentation)
headinclude,footinclude, % Extra spacing for the header and footer
BCOR5mm, % Binding correction
]{scrartcl}

\input{structure.tex} % Include the structure.tex file which specified the document structure and layout

\usepackage[a4paper, total={160mm,257mm}]{geometry}
\usepackage{wrapfig}

\title{\normalfont\spacedallcaps{Cheatsheets for Authentication and Key Agreements in\protect\\2G, 3G, 4G, and 5G}} % The article title

\author{\spacedlowsmallcaps{Prajwol Kumar Nakarmi}\\\footnotesize{\textit{Ericsson, Stockholm}}}

\date{2021} % An optional date to appear under the author(s)

\begin{document}

%----------------------------------------------------------------------------------------
%	HEADERS
%----------------------------------------------------------------------------------------

\renewcommand{\sectionmark}[1]{\markright{\spacedlowsmallcaps{#1}}} % The header for all pages (oneside) or for even pages (twoside)
\lehead{\mbox{\llap{\small\thepage\kern1em\color{halfgray} \vline}\color{halfgray}\hspace{0.5em}\rightmark\hfil}} % The header style

\pagestyle{scrheadings} % Enable the headers specified in this block

%----------------------------------------------------------------------------------------
%	TABLE OF CONTENTS & LISTS OF FIGURES AND TABLES
%----------------------------------------------------------------------------------------

\maketitle % Print the title/author/date block

\setcounter{tocdepth}{2} % Set the depth of the table of contents to show sections and subsections only

\tableofcontents % Print the table of contents

%\listoffigures % Print the list of figures

%----------------------------------------------------------------------------------------
%	ABSTRACT
%----------------------------------------------------------------------------------------

\section*{Abstract} % This section will not appear in the table of contents due to the star (\section*)

Authentication and Key Agreement (AKA) is a type of security protocol, used in 3GPP mobile networks, that provides two security capabilities. The first capability, called authentication, is to cryptographically assert that a mobile phone or a network is indeed who it claims to be, and the second capability, called key agreement, is to put necessary cryptographic keys in place for protection of traffic between the mobile phone and the network. Jointly, these two capabilities lay the foundation of secure 3GPP mobile networks. From 2G-5G, there are eight main versions of AKA, details of which are spread over and embedded deep in multiple technical specifications. It is getting increasingly difficult to quickly check a certain property of a certain AKA, let alone grasp the full picture of all AKAs. Therefore, I have prepared cheatsheets for all AKA versions in \S \ref{sec:cheatsheets} and listed their main properties in \S \ref{sec:aka}. I hope these will benefit university students, security researchers, and 3GPP standardization community. I welcome any corrections and feedback at \texttt{prajwol.kumar.nakarmi@ericsson.com}.

\newpage

%----------------------------------------------------------------------------------------
%	INTRODUCTION
%----------------------------------------------------------------------------------------

\section{3GPP mobile networks}
3GPP \cite{3gpp} has developed more than 10 releases of technical specifications for five generations of mobile networks, 2G-5G, over three decades. In 2018, the first set of 5G technical specifications was completed as 3GPP Release 15. With each release and especially with each new generation of mobile networks, additional features and enhancements are introduced to address contemporary needs and to enable novel use cases. 

\begin{figure}[h]
    \centering 
    \includegraphics[width=0.75\columnwidth]{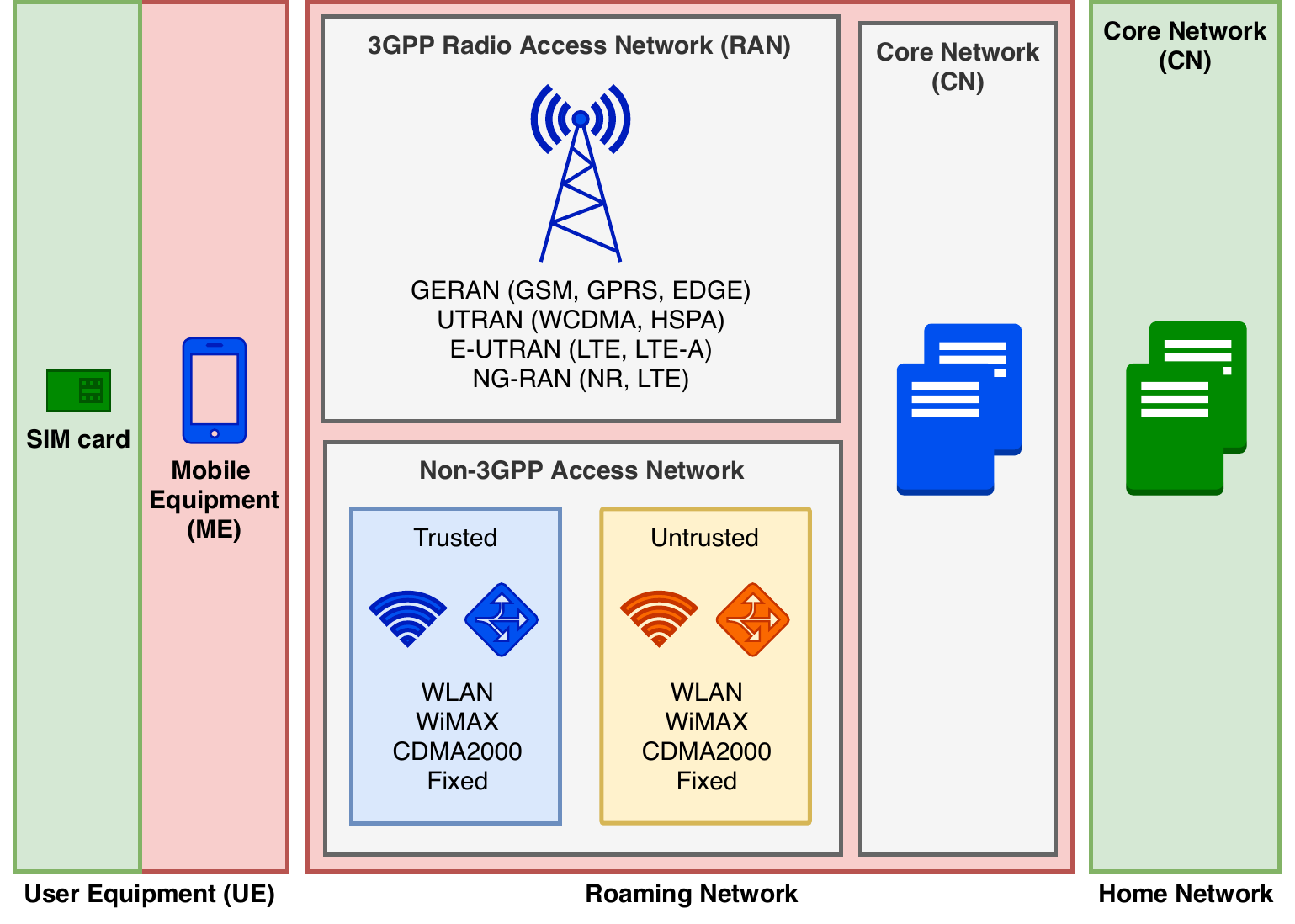} 
    \caption[A typical 3GPP mobile network]{A typical 3GPP mobile network, technical specifications for which are available at \cite{3gpp_specs}.} % The text in the square bracket is the caption for the list of figures while the text in the curly brackets is the figure caption
    \label{fig:3gpp_mobile_network} 
\end{figure}

A mobile network is maintained, and its services are offered by its operator called Mobile Network Operator (MNO). Fig. \ref{fig:3gpp_mobile_network} is a high-level overview of a 3GPP mobile network comprising of User Equipment (UE), Roaming network, and Home network. To use a network offered by a particular MNO, users are required to have a contractual relationship, called the subscription, with that MNO. In case of roaming, the Roaming network, where a user is being served, confirms the user's subscription by communicating with the Home network where the user has its subscription. The Roaming network is interchangeably called as visited or serving network. When the user is not roaming, both the Roaming and the Home network are the same and belong to the same MNO.

The UE consists of two parts, that are, SIM card and Mobile Equipment (ME). I use the popular term "SIM card" rather loosely to denote a smart card that provides a subscription. The term historically came from Subscriber Identity Module (SIM) that is a combined software and hardware module initially used in 2G. Since 3G, the software and hardware are separated into Universal Subscriber Identity Module (USIM) and Universal Integrated Circuit Card (UICC) respectively. The ME part, which is also sometimes referred to as Mobile Station (MS), represents what is popularly known as mobile phone, or smart phones. 

The access network is a part of the Roaming network and is responsible for providing communication to UEs by connecting them to the core network. It covers two broad categories, that are, 3GPP and non-3GPP access networks. The 3GPP access networks are the Radio Access Networks (RAN) specified in 3GPP, that are, 2G GERAN, 3G UTRAN, 4G E-UTRAN, and 5G NG-RAN. The non-3GPP access networks comprise of those accesses that are not specified in the 3GPP, including WLAN, WiMAX, CDMA2000, and fixed networks. Further, the non-3GPP accesses could be either trusted or untrusted. Note that being trusted or untrusted is not a characteristic of any particular non-3GPP access by itself. Rather, it is at the discretion of the Home network MNO.

The core network (CN) functionalities are distributed between the Roaming and Home networks. The functionalities include setting up security, handling UE mobilities, packet routing and forwarding, interconnecting to data network, and so on.

\newpage
%----------------------------------------------------------------------------------------
%	Authentication and key agreement
%----------------------------------------------------------------------------------------

\section{Authentication and key agreement (AKA) in 3GPP}
Security is one of the topics in which substantial improvements have been made from 2G to 5G. Today, 5G is the most secure and trustworthy mobile network. The foundation of its security lies in cryptography, particularly two fundamental processes, that are, authentication and key agreement.

\paragraph{Authentication.} I mentioned earlier that a user needs to have a subscription with an MNO in order to be able to access the MNO's network. Users with the MNO's subscription are allowed to use services offered by the MNO's network, for example, SMSes, phone calls, and Internet access. The MNOs in-turn bill or charge the users for the services they have used. For this business model to work, the network needs the user's unique long-term subscription identifier called International Mobile Subscriber Identity (IMSI) in 2G, 3G, and 4G, or Subscription Permanent Identifier (SUPI) in 5G. The UE provides this identifier stored in the SIM card to the network, thus, identifying the user. However, in order to ensure that the user cannot deny the bill or to prevent a fraudulent user from impersonating someone else's identifier, the network must assert that the user is who it claims to be. Authentication provides such assertion. 

In 3GPP, authentication is a security process that cryptographically asserts or proves that a given subscription identifier indeed belongs to a particular user. There is also another side of the authentication in which the UE asserts that the network is indeed who the network claims to be. Such assertion enables the UE to detect if an unauthorized party (such as an IMSI catcher) is trying to engage with the UE. When both the network and the UE authenticate each other, mutual authentication is achieved.

\paragraph{Key agreement.} While indispensable, authentication alone is not sufficient because the security of mobile networks requires secure transport of traffic as well. Other security capabilities, mainly confidentiality/ciphering, integrity protection, and replay protection of traffic are also necessary. Confidentiality/ciphering means encrypting the traffic to make it infeasible for unauthorized receivers to decrypt and read the original message. Integrity protection means adding a message authentication code to the traffic to make it infeasible for unauthorized parties to tamper the original message without the receiver detecting the tampering. Replay protection means keeping track of traffic to make it infeasible for unauthorized parties to re-send a previously valid traffic without the receiver detecting the replay. Cryptographic means of achieving these security capabilities require security keys. Key agreement is what provides the required security keys. 

In 3GPP, key agreement is a security process that enables the UE and the network to establish one or more shared security keys for protection of traffic sessions.

\paragraph{AKA.} The combination of above-mentioned two processes is intuitively called Authentication and Key Agreement, AKA.  It is a security protocol in a form of challenge-response protocol in which the network provides a cryptographic challenge, and the UE provides a cryptographic response. 

3GPP has always revised AKA for improvements when developing a new generation. One example of improvement on the authentication side is that while only the network authenticated the UE in 2G (one-sided authentication), the UE authenticated the network as well in later generations (mutual authentication). Similar example of the key agreement improvements is that while only encryption key was setup in 2G, additional key for integrity protection was also setup in later generations.  

There are eight main versions of AKA in 3GPP from 2G to 5G. What is common to them is that they are based on symmetric cryptography, a root pre-shared symmetric key (K) being shared between the mobile network and the SIM card. From this root K, security keys for protection of traffic sessions are derived and used between the ME and the mobile network. However, the root K itself is never exposed outside of a dedicated core network function in the mobile network and the SIM card. Note that both the SIM (2G) and the UICC (3G, 4G, 5G) are considered tamper resistant secure hardware component.

\section{Cheatsheets \label{sec:cheatsheets}}
Collection of cheatsheets are appended below, followed by \S \ref{sec:aka} on technical details. 
\includepdf[addtotoc={1, subsection, 1, {2G/GSM/GPRS/EDGE AKA}, fig:2g_aka},fitpaper, ]{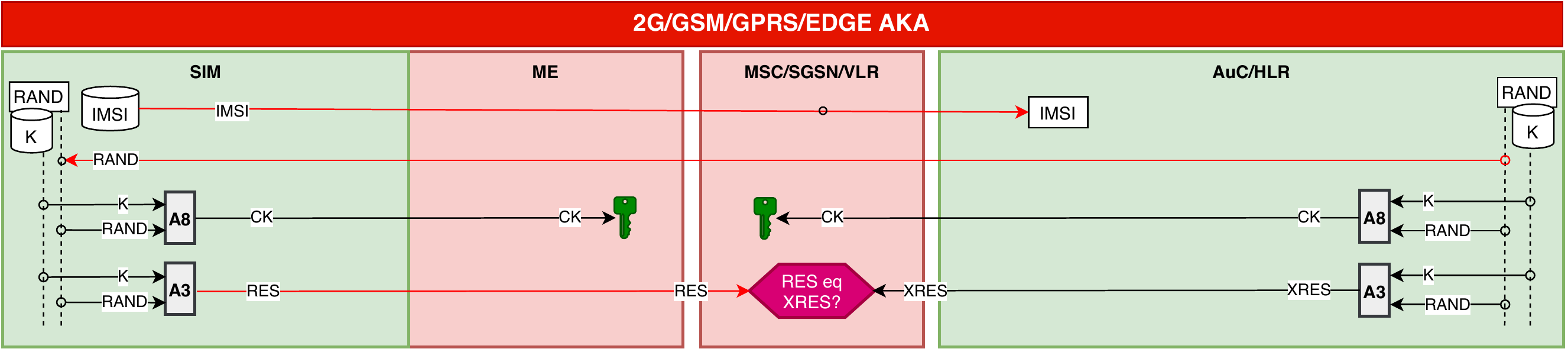}
\includepdf[addtotoc={1, subsection, 1, {3G/UMTS AKA (including EC-GSM-IoT)}, fig:3g_aka},fitpaper]{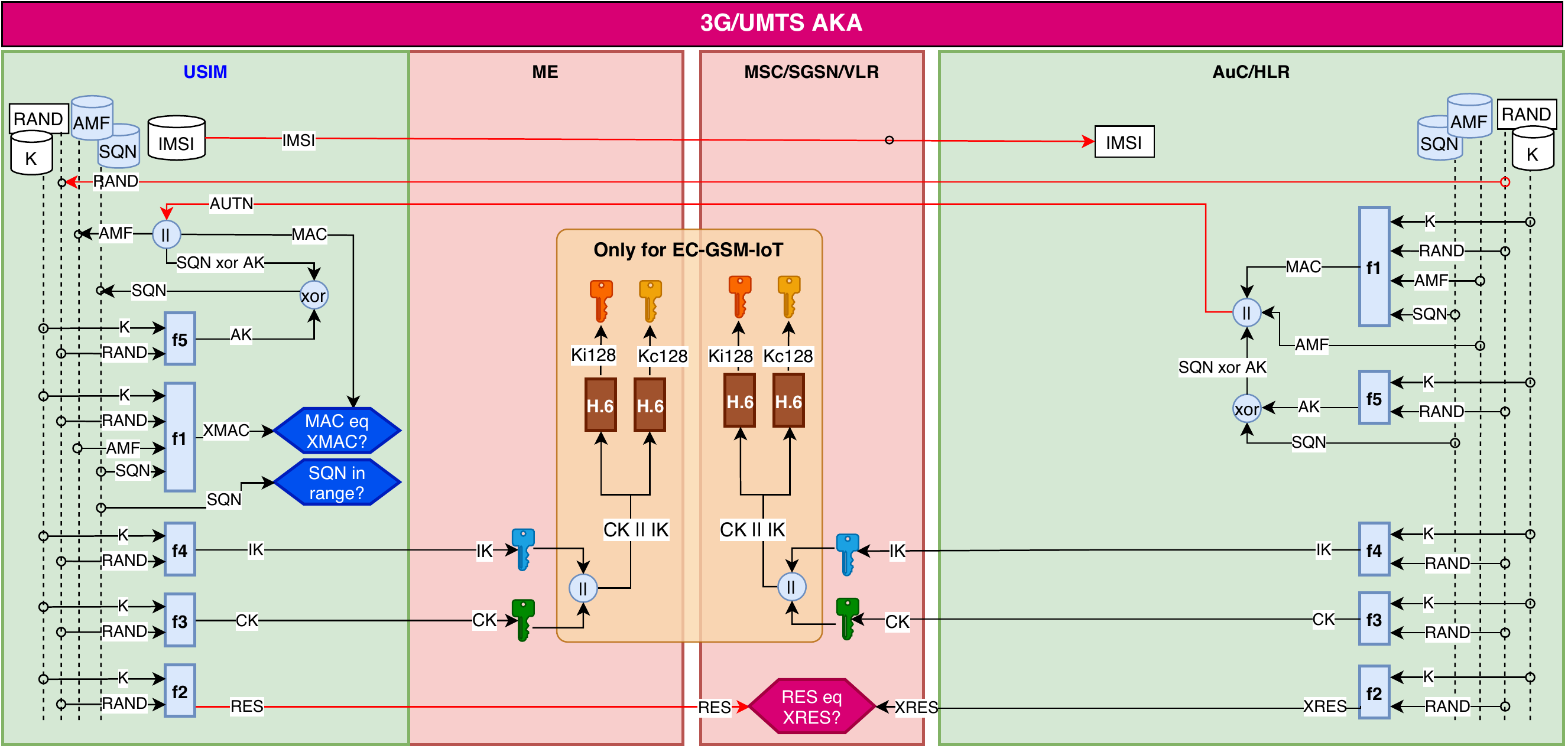}
\includepdf[addtotoc={1, subsection, 1, {4G/EPS AKA}, fig:4g_aka},fitpaper]{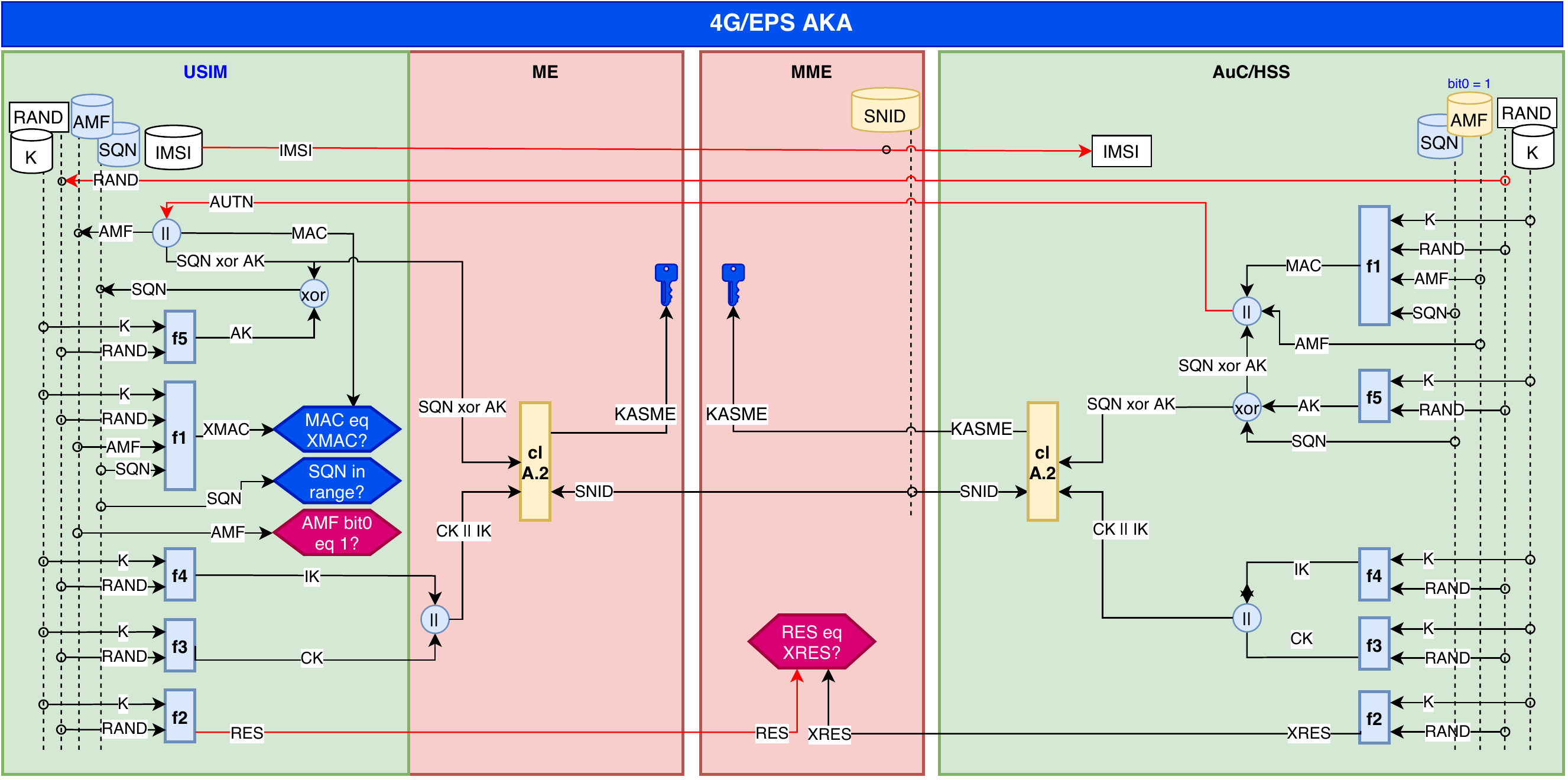}
\includepdf[addtotoc={1, subsection, 1, {4G/EPS EAP-AKA}, fig:4g_non-3gpp-eap-aka},fitpaper]{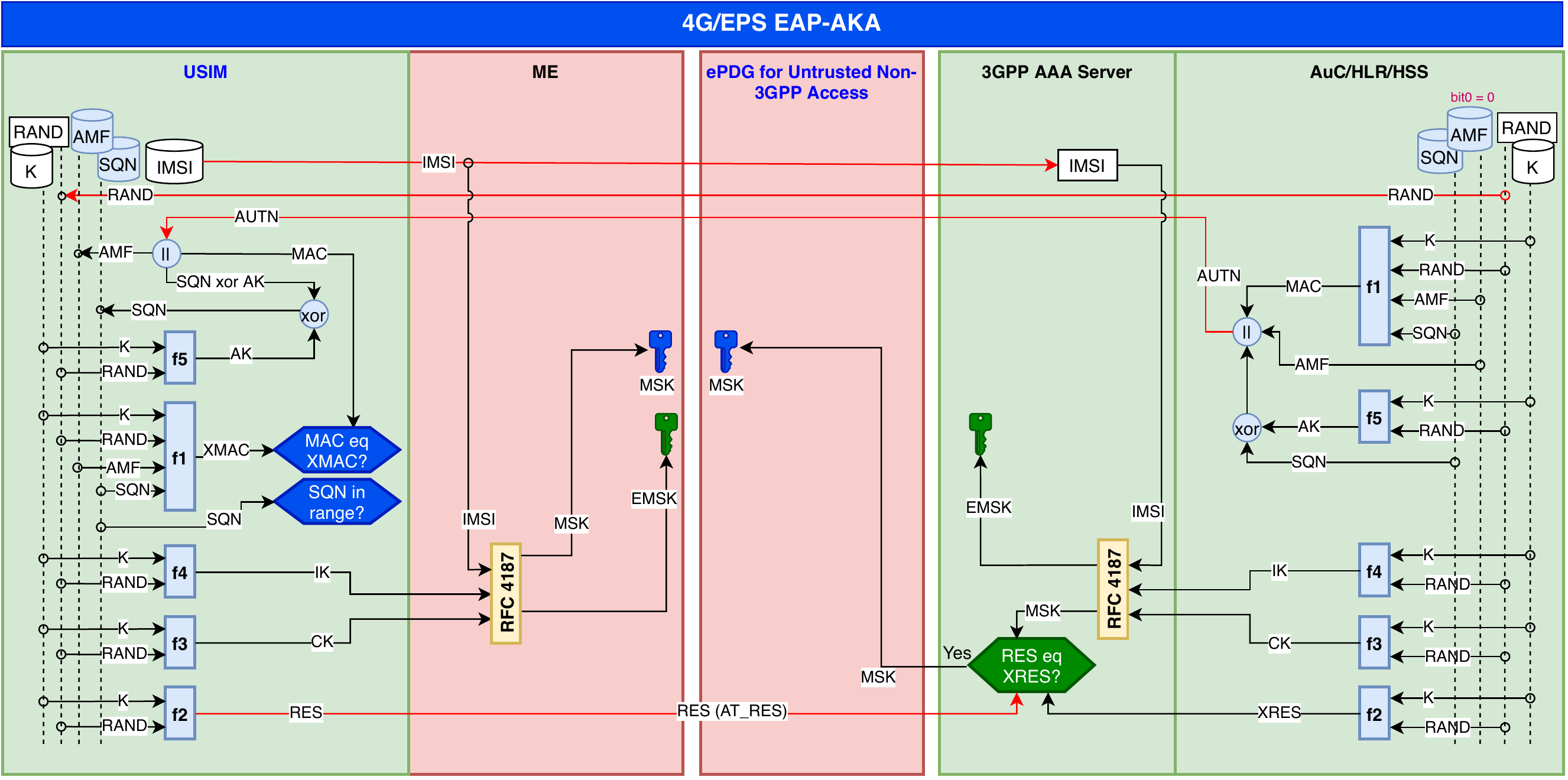}
\includepdf[addtotoc={1, subsection, 1, {4G/EPS EAP-AKA'}, fig:4g_non-3gpp-eap-aka-p},fitpaper]{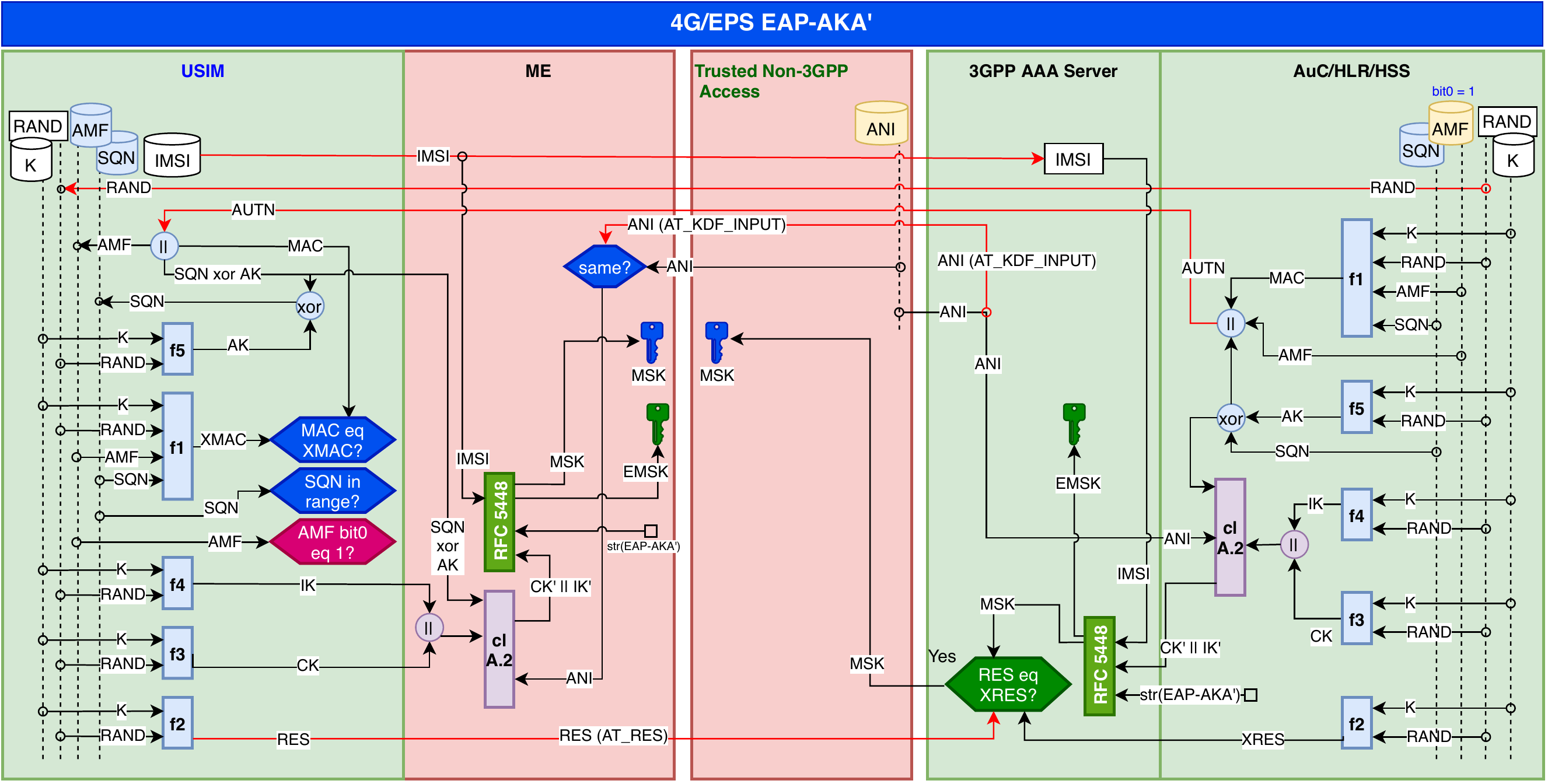}
\includepdf[addtotoc={1, subsection, 1, {5G AKA}, fig:5g_aka},fitpaper]{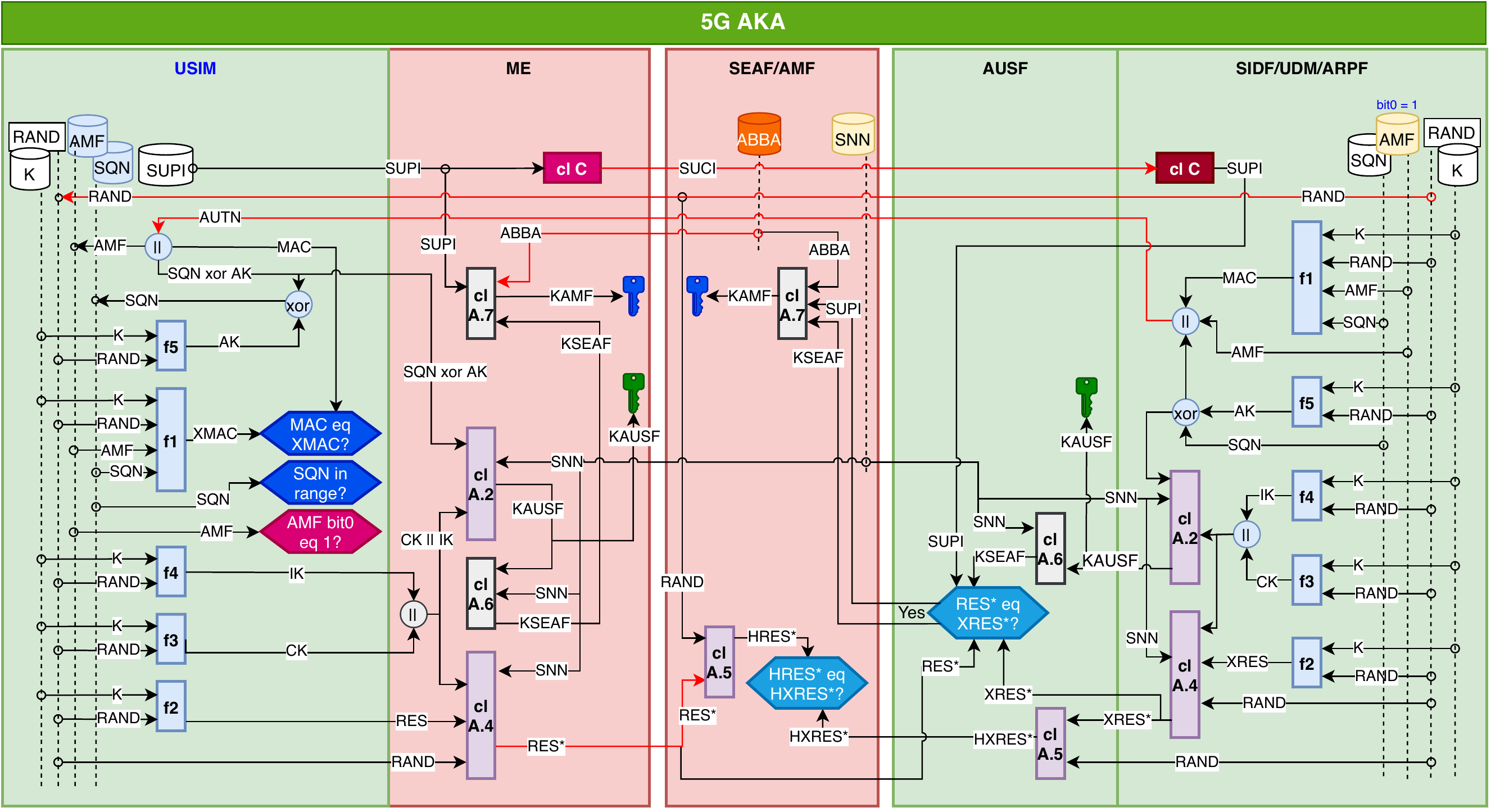}
\includepdf[addtotoc={1, subsection, 1, {5G EAP-AKA'}, fig:5g_eap-aka-p},fitpaper]{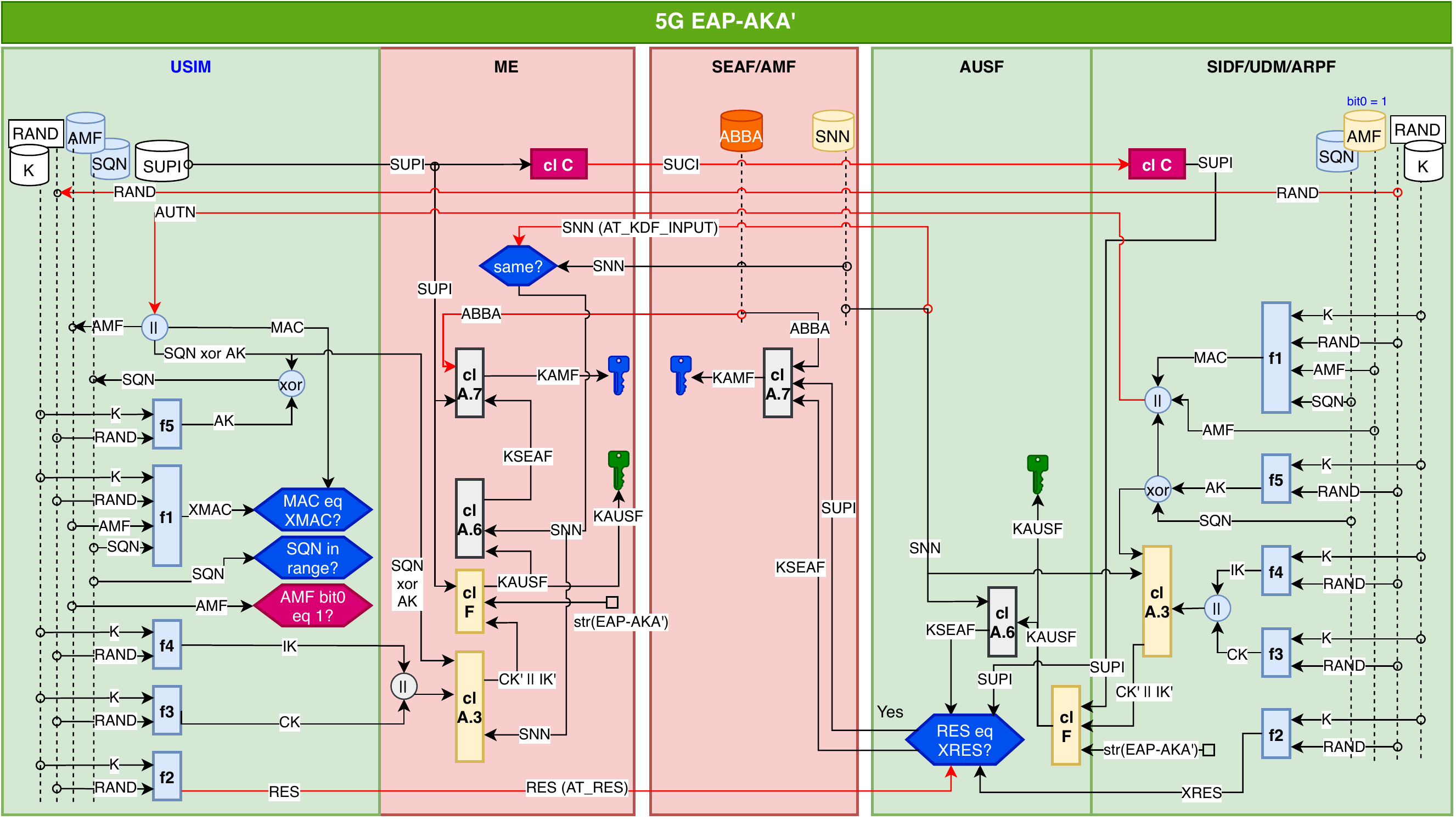}

\newpage
%----------------------------------------------------------------------------------------
%	AKA
%----------------------------------------------------------------------------------------
\section{Main properties of different AKA versions \label{sec:aka}}
%------------------------------------------------

\subsection{AKA in 2G}
Cheatsheet for AKA in 2G is in \S \ref{fig:2g_aka}. Details and references for 2G security, including AKA in 2G, can be found in 3GPP TS 43.020 \cite{3gpp_43.020}.

I loosely use the term 2G to denote mobile networks using circuit-switching (CS) and/or packet-switching (PS) architecture with GSM, GPRS, and EDGE. In \S \ref{fig:2g_aka}, the Mobile services Switching Centre (MSC), Serving GPRS Support Node (SGSN), and Visitor Location Register (VLR) represent the 2G Roaming network CN. The Authentication Centre (AuC) and Home Location Register (HLR) represent the 2G Home network CN. I have used the terms K, CK, ME, RES, and XRES for consistency, although 3GPP TS 43.020 \cite{3gpp_43.020} uses the terms Ki, Kc, MS, and SRES. The A3 is an authentication algorithm and A8 is a ciphering key generating algorithm. Annex C of 3GPP TS 43.020 \cite{3gpp_43.020} describes that these A3 and A8 algorithms are at the Home network MNO's discretion.
\\

\begin{tcolorbox}[colback=black!5!white,colframe=black!75!black,title=Main properties of AKA in 2G (cheatsheet in \S \ref{fig:2g_aka})]
\begin{itemize}[leftmargin=*]
    \item Authentication terminates at Roaming CN, although authentication vectors are provided by Home CN.
    \item The Roaming CN authenticates the UE by checking 32-bits RES and XRES. UE does not authenticate the network.
    \item The 64-bit CK is shared between the UE and the Roaming CN. This enables encryption but not integrity protection.
    \item Same AKA method is run separately for CS and PS.
\end{itemize}
\end{tcolorbox}

For UEs supporting enhanced GPRS in relation to Cellular IoT, called EC-GSM-IoT, 2G AKA is not used, rather 3G AKA as described in next section is used. 
\\

\subsection{AKA in 3G}
Cheatsheet for AKA in 3G is in \S \ref{fig:3g_aka}. 3G security, including AKA, is specified in 3GPP TS 33.102 \cite{3gpp_33.102}.

The term 3G or UMTS loosely denotes mobile networks using CS/PS with UTRAN. The f1 and f2 denote message authentication functions. In \S \ref{fig:3g_aka}, the f3, f4, and f5 denote key generating functions. These functions are at the Home network MNO's discretion. Information on MILENAGE and TUAK can be found in 3GPP 35-series specifications \cite{3gpp_35_series}. For UEs supporting EC-GSM-IoT, this AKA in 3G is used with some peculiarities, not the 2G AKA. The H.6 refers to Annex H.6 in 3GPP TS 43.020 \cite{3gpp_43.020}.
\\

\begin{tcolorbox}[colback=black!5!white,colframe=black!75!black,title=Main properties of AKA in 3G (cheatsheet in \S \ref{fig:3g_aka})]
\begin{itemize}[leftmargin=*]
    \item USIM is used, not SIM.
    \item Authentication terminates at Roaming CN (authentication vectors provided by Home CN).
    \item The Roaming CN authenticates the UE by checking 32- to 128-bits RES and XRES.
    \item The UE authenticates the Home CN by checking 64-bits MAC and XMAC and by verifying that 48-bits SQN is within acceptable range. Now, there is mutual authentication that enables the UE to determine if the network is legitimate (and not an IMSI catcher).
    \item The 128-bit CK and 128-bit IK are shared between the UE and the Roaming CN. This enables both encryption and integrity protection, further enhancing protection against IMSI catchers.
    \item Same AKA method is run separately for CS and PS.
\end{itemize}
\end{tcolorbox}

\begin{tcolorbox}[colback=black!5!white,colframe=black!75!black,title=Main properties of AKA in 3G used for EC-GSM-IoT (cheatsheet in \S \ref{fig:3g_aka})]
\begin{itemize}
    \item USIM is used, not SIM.
    \item Authentication terminates at Roaming CN (authentication vectors provided by Home CN).
    \item The Roaming CN authenticates the UE by checking 32- to 128-bits RES and XRES.
    \item The UE authenticates the Home CN by checking 64-bits MAC and XMAC and by verifying that 48-bits SQN is within acceptable range. There is mutual authentication.
    \item The 128-bit Kc128 and 128-bit Ki128 are shared between the UE and the Roaming CN, enabling both encryption and integrity protection.
\end{itemize}
\end{tcolorbox}

\subsection{AKA in 4G}
Cheatsheets for AKA in 4G are in \S \ref{fig:4g_aka}, \S \ref{fig:4g_non-3gpp-eap-aka}, and \S \ref{fig:4g_non-3gpp-eap-aka-p}. 4G security for 3GPP access, including AKA for 3GPP access, is specified in 3GPP TS 33.401 \cite{3gpp_33.401}. Similarly, 4G security for untrusted and trusted non-3GPP access, including AKA for them, is specified in 3GPP TS 33.402 \cite{3gpp_33.402}.

\bigskip
\begin{tcolorbox}[colback=black!5!white,colframe=black!75!black,title=Main properties of AKA in 4G for 3GPP access (cheatsheet in \S \ref{fig:4g_aka})]
    \begin{itemize}[leftmargin=*]
        \item USIM is used, not SIM.
        \item Authentication terminates at Roaming CN (authentication vectors provided by Home CN).
        \item The Roaming CN authenticates the UE by checking 32- to 128-bits RES and XRES.
        \item The UE authenticates the Home CN by checking 64-bits MAC and XMAC and by verifying that 48-bits SQN is within acceptable range. There is mutual authentication.
        \item The UE authenticates the Roaming CN by binding the 24-bit Serving Network Identity (SNID) to KASME. This enables serving network binding so that session keys generated for one Roaming network cannot be used in another.
        \item UE checks service type by verifying that Authentication Management Field (AMF) bit 0 is 1.
        \item The 256-bit KASME is shared between the UE and the Roaming CN. From this, encryption and integrity protection keys are derived.
    \end{itemize}
\end{tcolorbox}

\bigskip
\begin{tcolorbox}[colback=black!5!white,colframe=black!75!black,title=Main properties of AKA in 4G for untrusted non-3GPP access (cheatsheet in \S \ref{fig:4g_non-3gpp-eap-aka})]
    \begin{itemize}[leftmargin=*]
        \item USIM is used, not SIM.
        \item Authentication terminates at Home CN.
        \item It is based on Extensible Authentication Protocol (EAP)-AKA specified in IETF RFC 4187 \cite{rfc_4187}.
        \item The Home CN authenticates the UE by checking 32- to 128-bits RES and XRES.
        \item The UE authenticates the Home CN by checking 64-bits MAC and XMAC and by verifying that 48-bits SQN is within acceptable range. There is mutual authentication.
        \item 512-bit Master Session Key (MSK) is shared between the UE and the Roaming CN.
        \item 512-bit Extended Master Session Key (EMSK) is shared between the UE and the Home CN.
    \end{itemize}
    \end{tcolorbox}

\bigskip
\begin{tcolorbox}[colback=black!5!white,colframe=black!75!black,title=Main properties of AKA in 4G for trusted non-3GPP access (cheatsheet in \S \ref{fig:4g_non-3gpp-eap-aka-p})]
\begin{itemize}[leftmargin=*]
    \item USIM is used, not SIM. 
    \item Authentication terminates at Home CN.
    \item It is based on EAP-AKA' specified in IETF RFC 5448 \cite{rfc_5448}.
    \item The Home CN authenticates the UE by checking 32- to 128-bits RES and XRES.
    \item The UE authenticates the Home CN by checking 64-bits MAC and XMAC and by verifying that 48-bits SQN is within acceptable range. There is mutual authentication.
    \item The UE authenticates the Roaming CN by explicitly checking the Access Network Identity (ANI) sent by the Home CN and binding the variable length ANI to CK'||IK'. This ensures that session keys generated for one Roaming network cannot be used in another.
    \item UE checks service type by verifying that the AMF bit 0 is 1. Service type is also bound to CK'||IK' because ANI contains the service code.
    \item MSK and EMSK are bound to the authentication method because of "EAP-AKA'" input string.
    \item MSK and EMSK are also bound to IMSI.
    \item 512-bit MSK is shared between the UE and the Roaming CN.
    \item 512-bit EMSK is shared between the UE and the Home CN.
\end{itemize}
\end{tcolorbox}

\bigskip
By 4G, I loosely denote the mobile networks using Evolved Packet Core (EPC). The access network could be either 3GPP or non-3GPP; the non-3GPP access could further be either trusted or untrusted. There are separate AKAs for each of these accesses. 
In \S \ref{fig:4g_aka}, the Mobility Management Entity (MME) and the AuC/Home Subscriber Server (HSS) denote the Roaming and Home network CN respectively. The cl A.2 refers to Annex A.2 in 3GPP TS 33.401 \cite{3gpp_33.401}. In \S \ref{fig:4g_non-3gpp-eap-aka}, for untrusted non-3GPP access, the Evolved Packet Data Gateway (ePDG) and 3GPP AAA Authentication Authorization Accounting (AAA) server represent the Roaming and Home CN respectively. In \S \ref{fig:4g_non-3gpp-eap-aka-p}, for trusted non-3GPP access, the cl A.2 refers to Annex A.2 in 3GPP TS 33.402 \cite{3gpp_33.402}.

\bigskip
\bigskip
\subsection{AKA in 5G}
Cheatsheets for AKA in 5G are in \S \ref{fig:5g_aka} and \S \ref{fig:5g_eap-aka-p}. 5G security, including two AKAs, is specified in 3GPP TS 33.501 \cite{3gpp_33.501}. 

I loosely use 5G to denote the mobile networks using 5G Core (5GC) with the access network being either 3GPP or trusted/untrusted non-3GPP.

Two AKAs -- 5G AKA and 5G EAP-AKA' -- are specified, either of which can be used for all accesses. There is also yet another vendor specific EAP-5G which I ignore since it does not impact the core AKA. 

In \S \ref{fig:5g_aka} and \S \ref{fig:5g_eap-aka-p}, the Security Anchor Function (SEAF), and Access and Mobility Management Function (AMF) represents Roaming CN. The Authentication Server Function (AUSF), Subscription Identifier De-concealing Function (SIDF), Unified Data Management (UDM), Authentication credential Repository and Processing Function (ARPF) represent Home CN. The cl A.2-A.7, cl C, and cl F refer to those clauses in 3GPP TS 33.501 \cite{3gpp_33.501}. I have not shown the handling of AT\_MAC between ME and AUSF in 5G EAP-AKA' for simplicity.

\begin{tcolorbox}[colback=black!5!white,colframe=black!75!black,title=Main properties of 5G AKA (cheatsheet in \S \ref{fig:5g_aka})]
\begin{itemize}
    \item USIM is used, not SIM. UE identifies itself with Subscription Concealed Identifier (SUCI) which can only be de-concealed by the Home CN.
    \item Authentication terminates both at Roaming and Home CN.    
    \item The Roaming CN authenticates the UE by checking 128-bits HRES* and HXRES*.
    \item The Home CN authenticates the UE by checking 128-bits RES* and XRES*.
    \item The UE authenticates the Home CN by checking 64-bits MAC and XMAC and by verifying that 48-bits SQN is within acceptable range. There is mutual authentication.
    \item The UE authenticates the Roaming CN by binding the variable length Serving Network Name (SNN) to KAUSF, and KSEAF. The SNN is also bound to RES* and XRES*. This provides serving network binding.
    \item UE checks service type by verifying that the AMF bit 0 is 1. Service type is also bound to KAUSF, KSEAF, RES*, and XRES* because SNN contains the service code string "5G".
    \item Anti-Bidding down Between Architectures (ABBA) is achieved by binding the ABBA value to KAMF.
    \item User identifier binding is achieved by binding SUPI to KAMF.
    \item 256-bits KAMF is shared between the UE and the Roaming CN.
    \item 256-bits KAUSF is shared between the UE and the Home CN.
\end{itemize}
\end{tcolorbox}

\begin{tcolorbox}[colback=black!5!white,colframe=black!75!black,title=Main properties of 5G EAP-AKA' (cheatsheet in \S \ref{fig:5g_eap-aka-p})]
\begin{itemize}
    \item USIM is used, not SIM. UE identifies itself with Subscription Concealed Identifier (SUCI) which can only be de-concealed by the Home CN.
    \item Authentication terminates at Home CN.
    \item It is based on EAP-AKA' specified in IETF RFC 5448 \cite{rfc_5448} which is being updated in \cite{rfc_pfs}.    
    \item The Home CN authenticates the UE by checking 32- to 128-bits RES and XRES.
    \item The UE authenticates the Home CN by checking 64-bits MAC and XMAC and by verifying that 48-bits SQN is within acceptable range. There is mutual authentication.
    \item The UE authenticates the Roaming CN by explicitly checking the SNN sent by the Home CN and binding the variable length SNN to CK'||IK', KAUSF, and KSEAF. This ensures that session keys generated for one Roaming network cannot be used in another.
    \item UE checks service type by verifying that the AMF bit 0 is 1. Service type is also bound to CK'||IK', KAUSF, and KSEAF because the SNN contains the service code string "5G".
    \item KAUSF (and implicitly KSEAF too) is bound to the authentication method because of "EAP-AKA'" input as a string.
    \item Anti-Bidding down Between Architectures (ABBA) is achieved by binding the ABBA value to KAMF.
    \item User identifier binding is achieved by binding SUPI to KAMF.
    \item 256-bits KAMF is shared between the UE and the Roaming CN.
    \item 256-bits KAUSF is shared between the UE and the Home CN.
\end{itemize}
\end{tcolorbox}

\section{Summary}
Various revisions of AKA are definitely the showcase of how well 3GPP has addressed known and future threats to provide solid foundation of secure mobile networks. If I have to summarize by picking one major feature of AKA in each generation, that would be as below.

2G had only one-sided authentication in which the UE is authenticated by the network. But UE did not have means to authenticate the network which made it easier for IMSI catchers to trick the UE. 

This was fixed in 3G by introducing mutual authentication between the UE and the Home network. Still, it was possible that session keys generated for one Roaming network are valid for another Roaming network, meaning no cryptographic separation of security keys between Roaming networks. 

In 4G, this was fixed by cryptographically binding the Roaming network identifier with session keys. But, the authentication still terminated in the Roaming network which informs the Home network that a certain UE is in its network. The Home network had no means of guaranteeing if the UE was really present in that Roaming network. 

To fix this, authentication in 5G is terminated in the Home network. So far so good. Any more enhancements to come about in 6G? -- I'm on it.

%----------------------------------------------------------------------------------------
%	BIBLIOGRAPHY
%----------------------------------------------------------------------------------------

\renewcommand{\refname}{\spacedlowsmallcaps{References}} % For modifying the bibliography heading

\bibliographystyle{unsrt}

\bibliography{cheatsheet_aka_2g_3g_4g_5g.bib} % The file containing the bibliography

%----------------------------------------------------------------------------------------

\end{document}

%% file: structure.tex
\usepackage[
nochapters, % Turn off chapters since this is an article        
beramono, % Use the Bera Mono font for monospaced text (\texttt)
eulermath,% Use the Euler font for mathematics
pdfspacing, % Makes use of pdftex’ letter spacing capabilities via the microtype package
dottedtoc % Dotted lines leading to the page numbers in the table of contents
]{classicthesis} % The layout is based on the Classic Thesis style

\usepackage{arsclassica} % Modifies the Classic Thesis package

\usepackage[T1]{fontenc} % Use 8-bit encoding that has 256 glyphs

\usepackage[utf8]{inputenc} % Required for including letters with accents

\usepackage{graphicx} % Required for including images
\graphicspath{{Figures/}} % Set the default folder for images

\usepackage{enumitem} % Required for manipulating the whitespace between and within lists

\usepackage{lipsum} % Used for inserting dummy 'Lorem ipsum' text into the template

\usepackage{subfig} % Required for creating figures with multiple parts (subfigures)

\usepackage{amsmath,amssymb,amsthm} % For including math equations, theorems, symbols, etc

\usepackage{varioref} % More descriptive referencing

\usepackage{pdfpages}
\usepackage{tcolorbox}

\theoremstyle{definition} % Define theorem styles here based on the definition style (used for definitions and examples)

\theoremstyle{plain} % Define theorem styles here based on the plain style (used for theorems, lemmas, propositions)

\theoremstyle{remark} % Define theorem styles here based on the remark style (used for remarks and notes)

\hypersetup{
colorlinks=true, breaklinks=true, bookmarks=true,bookmarksnumbered,
urlcolor=webbrown, linkcolor=RoyalBlue, citecolor=webgreen, % Link colors
pdftitle={}, % PDF title
pdfauthor={\textcopyright}, % PDF Author
pdfsubject={}, % PDF Subject
pdfkeywords={}, % PDF Keywords
pdfcreator={pdfLaTeX}, % PDF Creator
pdfproducer={LaTeX with hyperref and ClassicThesis} % PDF producer
}